\documentclass[twocolumn,floatfix,pra,showpacs,aps,superscriptaddress]{revtex4-1}
   
\usepackage{amsmath}
\usepackage{amsbsy}
\usepackage{amsthm}
\usepackage{amssymb}
\usepackage{latexsym}
\usepackage[dvips]{color,graphicx}

\begin{document}

\title{Expansion of Bose-Hubbard Mott insulators in optical lattices}

\author{Mark Jreissaty}
\affiliation{Department of Physics, Georgetown University, Washington, DC 20057, USA}

\author{Juan Carrasquilla}
\affiliation{Department of Physics, Georgetown University, Washington, DC 20057, USA}

\author{F. Alexander Wolf}
\affiliation{Department of Physics, Georgetown University, Washington, DC 20057, USA}
\affiliation{Theoretical Physics III, Center for Electronic Correlations and Magnetism,
Institute of Physics, Augsburg University, D-86135 Augsburg, Germany}

\author{Marcos Rigol}
\affiliation{Department of Physics, Georgetown University, Washington, DC 20057, USA}

\begin{abstract}

We study the expansion of bosonic Mott insulators in the presence of an optical lattice
after switching off a confining potential. We use the Gutzwiller mean-field approximation 
and consider two different setups. In the first one, the expansion is restricted to 
one direction. We show that this leads to the emergence of two condensates with well 
defined momenta, and argue that such a construct can be used to create atom lasers in optical 
lattices. In the second setup, we study Mott insulators that are allowed 
to expand in all directions in the lattice. In this case, a simple condensate is seen to 
develop within the mean-field approximation. However, its constituent bosons are found to 
populate many nonzero momentum modes. An analytic understanding of both phenomena in terms 
of the exact dispersion relation in the hard-core limit is presented.
\end{abstract}
\pacs{03.75.Kk,03.75.-b,03.75.Hh,05.30.Jp}
\maketitle

\section{Introduction}

In recent years, the study of strongly correlated ultracold gases in optical 
lattices has become an important focus of experimental and theoretical research
\cite{bloch_dalibard_review_08,cazalilla_citro_11}. Initially, many of the efforts 
were devoted to studying equilibrium quantum phase transitions, for which the 
observation of the superfluid to Mott-insulator phase transitions in three 
\cite{greiner_mandel_02a}, two \cite{spielman_phillips_07}, and one \cite{stoferle_moritz_04} 
dimension was a major experimental accomplishment. More recently, the study of the 
nonequilibrium dynamics of these systems has started gaining attention 
\cite{greiner_mandel_02b,kinoshita_wenger_06,hofferberth_lesanovsky_07,hung_zhang_10,trotzky_chen_11}.
The latter is possible, and offers great insights into the coherent dynamics 
of quantum systems, thanks to the nearly ideal isolation of the gas from the 
environment, and to the unique control that has been achieved experimentally over 
the parameters that determine the dynamics.

The nonequilibrium dynamics of quantum systems is a rich subject; see, e.g., 
Refs.~\cite{dziarmaga_10,cazalilla_rigol_10,polkovnikov_sengupta_11}. In this
paper, we will be mainly interested in the expansion of a strongly interacting bosonic 
gas in the presence of an optical lattice. 
Experimentally, expansions (time-of-flight measurements) have been used to learn about 
the properties of equilibrium gases \cite{dalfovo_giorgini_review_99}. The idea behind 
those measurements is that, if the gas is allowed to expand in the absence of interactions, 
the initial momentum distribution function fully determines the density distribution after 
a long expansion time. Hence, by taking a picture of the latter, one can determine the 
former by means of a simple rescaling 
\cite{dalfovo_giorgini_review_99,bloch_dalibard_review_08,cazalilla_citro_11}. In such 
experiments, the particles are allowed to expand in absence of both the trapping potential 
and the underlying optical lattice. This implies that the kinetic energy is significantly larger 
than the interaction energy and, therefore, interactions can generally be neglected during 
the expansion. On the other hand, in the presence of the optical lattice where interactions cannot 
be neglected during the expansion, unusual phenomena can occur \cite{sutherland_98},  e.g., 
the transformation of the momentum distribution of an expanding gas of impenetrable bosons into 
the momentum distribution function of a noninteracting Fermi gas in equilibrium, which has a 
Fermi edge \cite{rigol_muramatsu_05eHCBb,minguzzi_gangardt_05}. In addition, it has been recently 
proposed that the expansion of two-component fermions in the presence of strong interactions 
in an optical lattice can be used as a tool to cool the gas through a ``quantum distillation''
 process \cite{fabian_manmana_09}.

There is another remarkable phenomenon that occurs during the expansion
of a strongly interacting gas of bosons in an optical lattice, and that is the emergence
of coherence during the expansion of Mott-insulating states. This phenomenon 
will be the main focus of the present paper. A Mott insulator
is an interaction-generated insulator that exhibits a gap to one-particle
excitations. The presence of this gap produces an exponential decay of the 
one-particle correlations. Within the Bose-Hubbard
model description, the Mott insulator is the ground state of lattice bosons for commensurate
fillings and sufficiently strong interactions \cite{fisher_weichman_89}. As mentioned 
before, Mott insulators have been created experimentally using ultracold gases in optical 
lattices. Due to the presence of a trapping potential, the Mott 
insulating phases in experiments usually coexist with superfluid domains. 
Surprisingly, when Mott insulators are allowed to expand by turning off the trap
in the presence of a lattice, coherence emerges between initially uncorrelated 
particles \cite{rigol_muramatsu_04eHBCa,rigol_muramatsu_05eHCBc,rodriguez_manmana_06}, 
a phenomenon that resembles a sort of dynamical phase transition.

The study in Ref.~\cite{rigol_muramatsu_04eHBCa} demonstrated that the expansion
of pure Mott-insulating states of hard-core bosons in one dimension (1D) leads to the 
development of quasi-long-range correlations and to the emergence of quasi-condensates at 
finite momenta $k=\pm \pi/2a$ (where $a$ is the lattice parameter). In a 
related setup, the onset of quasi-long-range correlations has been 
studied more recently (in the spin-1/2 language) using various analytic 
and approximate approaches \cite{lancaster_mitra_10}. It is important to
note that these results were obtained in the limit of infinite on-site repulsions 
between the bosons (hard-core limit), and for pure Mott-insulating domains.
It is, therefore, natural to wonder whether the same results can be observed
experimentally with ultracold gases, for which the on-site interactions
are not infinite and the insulating domains are always surrounded by 
superfluid ones. Systems containing hard-core bosons, in which
superfluid and insulating domains coexist, were studied in 
Ref.~\cite{rigol_muramatsu_05eHCBc}. It was found there that, if most of the 
initial system is in a Mott-insulating state, quasicondensates with momenta
$k\approx\pm \pi/2a$ are generated. The superfluid wings surrounding
the insulating domain in the initial density profiles appeared
to have no negative effect on the emergence of quasicoherence during the  
expansion. Furthermore, the expansion of a Mott insulator in a system with 
finite on-site interactions (soft-core bosons) was studied in 
Ref.~\cite{rodriguez_manmana_06}. Once again, quasi-long-range coherence 
developed during the expansion and lead to quasicondensates with momenta 
$|k|\lesssim \pi/2a$. The differences between the position of the peaks for 
soft-core bosons and hard-core bosons can be understood by considering their 
respective dispersion relations. Finally, a similar 
onset of power-law decaying correlations has been observed during the expansion of a Mott 
insulator within the fermionic Hubbard model \cite{fabian_rigol_08}.

A more challenging question is what happens in higher dimensions. This
question poses greater conceptual and computational challenges. Since
hard-core bosons in 1D are described by an integrable model, one may expect 
their behavior to be unique. Hence it may not extend to nonintegrable 
models far from any integrable limit, such as the Bose-Hubbard model 
with finite on-site interaction in higher dimensions. In the particular case 
of one spatial dimension, since the boson's on-site interaction has to be 
strong enough for a Mott insulator to exist, the results obtained for the 1D 
expansion \cite{rodriguez_manmana_06} can be thought as still affected by 
the infinite-repulsion integrable limit, and thus may not generalize to 
higher dimensions. Furthermore, while the 1D expansion can be studied by 
means of exact analytical and/or numerical approaches in the hard-core limit
\cite{rigol_muramatsu_04eHBCa,rigol_muramatsu_05eHCBc,lancaster_mitra_10} 
and by means of an unbiased numerical technique for the soft-core
case (the time-dependent-density-matrix-renormalization-group approach 
\cite{schollwock_review_05}), no such tool exists to study the expansion
in higher dimensions. For the latter, no efficient and unbiased, analytical,
or computational approach is known that can deal with the dynamics in
the presence of strong interactions. Approximations are therefore required
to study these systems.

A first step toward understanding what happens in higher dimensions
was taken in Ref.~\cite{hen_rigol_10}. In that work, which was a study 
in the hard-core limit utilizing the Gutzwiller mean-field approximation,
a three-dimensional (3D) Mott insulator was allowed to expand in a single 
direction in the lattice. The study showed that condensates emerge at finite 
momenta following the ``melting" of a Mott insulator. In addition,
the momenta of the resulting condensates were found to be fully determined
by the ratio between the hopping amplitudes transverse to and along the
expansion. Hence these systems can be used to create highly controllable
atom lasers. Other nonequilibrium setups that allow one to obtain condensation 
at finite momentum \cite{Smith1,Smith2} do not seem suitable for the 
latter purpose.

In this paper, we extend the results reported in Ref.~\cite{hen_rigol_10} 
to the expansion of Mott insulators with finite on-site interaction in 
two dimensions (2D). Our results in 2D can be straightforwardly extended 
to three dimensions. We first consider a setup similar to that in 
Ref.~\cite{hen_rigol_10}, in which soft-core bosons are allowed to 
expand in a single direction. We show that similarly to
the case in the hard-core limit, condensates emerge at finite momenta.
We then proceed to study a system in which the Mott insulator is
allowed to expand in all directions of the lattice. For this particular
case, we show that, at least within the mean-field approximation, 
a simple condensate with many different momenta develops as the Mott 
insulator expands. We also discuss how these results can be understood 
in terms of the exact dispersion relation in the hard-core limit.

The paper is organized as follows. In Sec.\ \ref{methods}, we introduce
the model and the time-dependent mean-field approach used to study
the expansion. The explanation of how our initial states are prepared 
is presented in Sec.~\ref{sec:inistate}. Section~\ref{laser} is
devoted to the study of the expansion along one direction. 
The case in which the Mott insulator is allowed to 
expand in all directions is discussed in Sec.~\ref{isotropic}. Finally, the 
conclusions are presented in Sec.~\ref{conclusions}.

\section{Model and mean-field approach} \label{methods}

Our study is relevant to ultracold bosons trapped in deep optical lattices. 
They can be well described by the Bose-Hubbard Hamiltonian \cite{jaksch_bruder_98}
\begin{eqnarray}
\hat{H} &=& -\sum_{\langle i,j \rangle}J_{ij} \left(\hat{a}^\dagger_{i} \hat{a}^{}_{j} + \textrm{H.c.} \right) 
+ \dfrac{U}{2} \sum_i \hat{n}_{i} \left( \hat{n}_{i} -1 \right) 
\nonumber \\ &&+ \sum_{i} (V_x\, x_i^2+V_y\, y_i^2)\, \hat{n}_{i} \,\,,
\label{HubbB}
\end{eqnarray}
where $\hat{a}^\dagger_{i}$ ($\hat{a}_{i}$) is the boson creation (annihilation) 
operator at a given site $i$, and $\hat{n}_{i} = \hat{a}^\dagger_{i} \hat{a}^{}_{i}$ 
is the particle number operator. The first two terms of the
expression are the usual kinetic and interaction terms that define the Hubbard
model in the homogeneous case, where $J_{ij}$ is the hopping amplitude between neighboring 
sites $i$ and $j$ ($\langle i,j \rangle$), and $U$ is the on-site
interaction strength. In experiments involving ultracold gases, a trap is 
required to confine the atoms. To a good approximation, those traps are 
harmonic, as considered in the last term in Eq.~\eqref{HubbB}. Here, $V_x$ and $V_y$ 
denote the strength of the trap in the $x$ and $y$ direction, while $x_i$ and $y_i$ are
the coordinates of site $i$ with respect to the center of the trap.
In the remainder of the manuscript, positions will be given in units of the lattice 
spacing $a$, which is taken to be the same in all directions.

The ground-state phase diagram of the homogeneous case ($V_x=V_y=0$ and $J_{ij}=J$) 
has been studied extensively in all dimensions
\cite{fisher_weichman_89,batrouni_scalettar_90,freericks_monien_96,
elstner_monien_99,kuhner_white_00,sansone_prokofev_07,sansone_soyler_08}.
As mentioned earlier, it contains both superfluid
and Mott-insulating phases. The former occurs for incommensurate
fillings, and for commensurate fillings for $U/J$ is below some 
critical value that depends on the filling and dimensionality of the system. 
The insulating phases are only present for commensurate fillings and sufficiently 
strong interactions. This model is not exactly solvable 
in any dimension. Unbiased results for the phase diagram have been obtained 
using quantum Monte Carlo simulations
\cite{batrouni_scalettar_90,sansone_prokofev_07,sansone_soyler_08},
density-matrix renormalization group (1D) \cite{kuhner_white_00}, and strong
coupling expansions \cite{freericks_monien_96,elstner_monien_99}.
Interestingly, a simple approximation such as the Gutzwiller mean-field theory, 
which is exact only in the limit of infinite dimensions, can still describe 
qualitatively the complete phase diagram of this model for finite 
dimensions \cite{fisher_weichman_89}.

In the presence of a trap, superfluid and Mott-insulating phases
coexist and form a ``wedding cake" structure 
\cite{batrouni_rousseau_02,wessel_alet_04,rigol_batrouni_09}.
The particular shape of the density distribution function can be understood
within the local density approximation \cite{bloch_dalibard_review_08}, given
the phase diagram in the homogeneous case (see Fig.~\ref{fig:LDA}, and the 
accompanying explanation). Experimentally, such a structure
has been observed by several groups
\cite{folling_widera_06,campbell_mun_06,gemelke_Zhang_09,bakr_peng_10,sherson_weitenberg_10}.
As mentioned in the Introduction, the unique opportunities provided by the study of 
these systems come from the experimentalist's high degree of control over
the depth and structure of the optical lattice and the confining potential. Essentially, 
all the microscopic parameters in Eq.~\eqref{HubbB} can be manipulated and made
time dependent. For example, $J$, $U$, and $V_x,V_y$ can be modified by changing 
the intensity of the laser beams that produce the lattice; $V_x,V_y$ can be modified 
by introducing an additional trapping (or antitrapping) magnetic field; and $U$ 
can be modified using Feshbach resonances. Together with the high degree
of isolation from the environment, this level of control makes these experiments
ideal to study the effects of strong correlations and dimensionality in the
nonequilibrium dynamics of the expanding gases.

In this paper, we are interested in studying the expansion of Mott-insulating 
states in the lattice. For this, we prepare the system in its ground state, for a 
particular number of particles, $U$, $J$, and $V_x,V_y$. At time $t=0$,
we turn off the trap and allow the system to expand  in the presence of the 
optical lattice. During the expansion, particles interact strongly and we compute 
observables such as the density, momentum distribution function, and condensate fraction, 
as a function of time. All our calculations are performed utilizing the time-dependent 
Gutzwiller mean-field theory. This is a good approximation in high dimensions and, 
in Ref.~\cite{hen_rigol_10}, it was shown to qualitatively reproduce the behavior
seen in an exact diagonalization study of the expansion in (small) 2D lattices.

This mean-field theory is based on the restriction that the wave function is of the 
Gutzwiller-type product state,
\begin{equation}\label{productState}
 \vert \Psi_{\mathrm{MF}} \rangle 
 = \prod_{i=1}^{L} \sum_{n=0}^{n_\text{c}} \gamma_{in} \vert n \rangle_i\, ,
 \end{equation}
where $L$ is the number of lattice sites, $n_c$ is a cutoff taken to be 
large enough such that all our results are independent of $n_c$ 
($n_c=3$ in our calculations), $\vert n \rangle_i$ are the Fock states for 
lattice site $i$, and the complex coefficients $\gamma_{in}$ are variational 
parameters determined by minimizing the expectation value
\begin{equation}\label{EQMin}
\langle \Psi_{\mathrm{MF}} \vert \hat{H}  - \mu \hat{N} \vert \Psi_{\mathrm{MF}} \rangle.
\end{equation}
In Eq.~\eqref{EQMin}, $\mu$ is the chemical potential, i.e., we work on the 
grand canonical ensemble,  and $\hat{N}$ the total number of particle operator.

The energy minimization leads to the set of equations
\begin{align}
&\, \sum_{\langle j \rangle_i}-J_{i j} \left[ \sqrt{n+1}\,\gamma_{i (n+1)}  \Phi_{j}^* 
+ \sqrt{n}\,\gamma_{i (n-1)}  \Phi_{j}\right] 
\nonumber \\ &\qquad\qquad+ n \left[\dfrac{U}{2}(n+1)+V_i-\mu\right]\ \gamma_{i n}=0,
\label{mfr0}
\end{align}
where $\sum_{\, \langle j \rangle_i}$ denotes a summation over all $j$ that are nearest neighbors of 
$i$, $V_i=V_x\, x_i^2+V_y\, y_i^2$, and $\gamma_{i(-1)} = \gamma_{i(n_\text{c} + 1)} = 0$. 
The mean-field $\Phi_j$ is defined as
\begin{align}
\Phi_j=\langle \hat{a}_j\rangle=\sum_{n=1}^{n_\text{c}}\sqrt{n}\,\gamma_{j(n-1)}^*\gamma_{jn}.
\label{mfpotential}
\end{align}
By solving Eq.~\eqref{mfr0}, one obtains all the coefficients $\gamma_{in}$ 
in Eq.~\eqref{productState}, and hence determines the initial state for the time 
evolution.

An alternative approach, which is equivalent to that of solving Eq.~\eqref{mfr0}, can be used.  
It consists in finding the ground state of the following mean-field decoupled 
Hamiltonian \cite{sheshadri_Krishnamurthy_93}: 
\begin{align}\label{meanfbose1}
\hat{{\cal H}}_{\mathrm{MF}}=- \sum_{\langle i,j \rangle} J_{i j}
\left( {\hat a}^\dagger_i \Phi_j + \Phi_i^* {\hat a}_j-\Phi_i^*\Phi_j \right) 
+ \textrm{H.c.} 
\nonumber \\
+ \frac{U}{2} \sum_i {\hat n}_i ({\hat n}_i-1) + 
\sum_i \left( V_i -\mu\right) {\hat n}_i.
\end{align}
This Hamiltonian is obtained by a decoupling of the hopping terms in Eq.~\eqref{HubbB} 
as 
\begin{equation}\label{decop}
{\hat a}^\dagger_i {\hat a}_j \simeq {\hat a}^\dagger_i  \Phi_j+{\hat a}_j \Phi_i^* - \Phi_i^* \Phi_j.
\end{equation}
Equation~\eqref{HubbB} can then be written as a sum over single-site Hamiltonians 
$\hat{{\cal H}}_{i}$, coupled only through constant neighboring terms $\Phi_j$. This 
implies that the ground state of the decoupled Hamiltonian can be written as the 
product state in Eq.~\eqref{productState}. The optimal set of coefficients 
$\gamma_{in}$ is related to the ground-state eigenvector components of the 
Hamiltonian in Eq.~\eqref{meanfbose1} subject to the condition in 
Eq.~\eqref{mfpotential}, which has to be self-consistently satisfied. In practice, 
this condition is reached as follows: an arbitrary  set of nonzero $\Phi_i^0$ is used 
to construct a set of decoupled Hamiltonians $\hat{{\cal H}}_{i}$ which are diagonalized for 
each site in order to obtain a set of $\gamma_{in}$'s. Given $\gamma_{in}$, a set of 
$\Phi_i^\textrm{new}$ is computed according to Eq.~\eqref{mfpotential}. If the elements on 
this new set are equal to those of the previous set $\Phi_i^0$  within a certain desired 
precision, then self-consistency has been reached. If not, we set 
$\Phi_i^0=\Phi_i^\textrm{new}$ and construct and a new set of decoupled Hamiltonians and 
the process of diagonalizing and finding $\Phi_i^\textrm{new}$ is iteratively repeated until 
self-consistency is satisfied. We have verified that both approaches based on 
Eq.~\eqref{mfr0} and Eq.~\eqref{meanfbose1} provide quantitatively consistent results. However,
at least within our implementations, the one based on the mean-field decoupled Hamiltonian 
is more stable and considerably faster than the one based on the direct solution of Eq.~\eqref{mfr0}. 

After turning off the trap, the time evolution of 
$|\psi_{\mathrm{MF}} \rangle$ is computed using the time-dependent variational 
principle \cite{amico_penna_98,jaksch_venturi_02,zakrzewski_01,snoek}, which minimizes 
the expectation value,
\begin{equation}\label{NEQMin}
\langle \Psi_{\mathrm{MF}} \vert  i \partial_t - \hat{H}  + \mu \hat{N} \vert \Psi_{\mathrm{MF}} \rangle \,,
\end{equation}
and leads to the following set of differential equations:
\begin{multline}\label{jakschEq}
 i \dot\gamma_{in} 
 = - \sum_{ \, \langle j \rangle_i} J_{ij}
\left[ \sqrt{n+1}\, \gamma_{i(n+1)} \Phi_j^* + \sqrt{n}\, \gamma_{i(n-1)} \Phi_j \right] \\
 + n \left[\, \frac{U}{2}(n-1) + V_i - \mu  \right]\gamma_{i n} \,.
\end{multline}
Here, we use the same notation as in Eq.~\eqref{mfr0}. 
It is important to note that Eq.~\eqref{jakschEq} determines the time evolution while 
preserving the normalization of the wave function and the total number of particles $N_b$
in the system. We solve this set of $L\times\left( n_\text{c}+1\right)$ differential equations numerically, 
using a fourth-order Runge-Kutta method. Self-consistency is enforced by monitoring the 
total energy, particle number, and the normalization.

\section{The initial state}\label{sec:inistate}

\begin{figure}
\includegraphics*[width=0.35\textwidth,angle=-90]{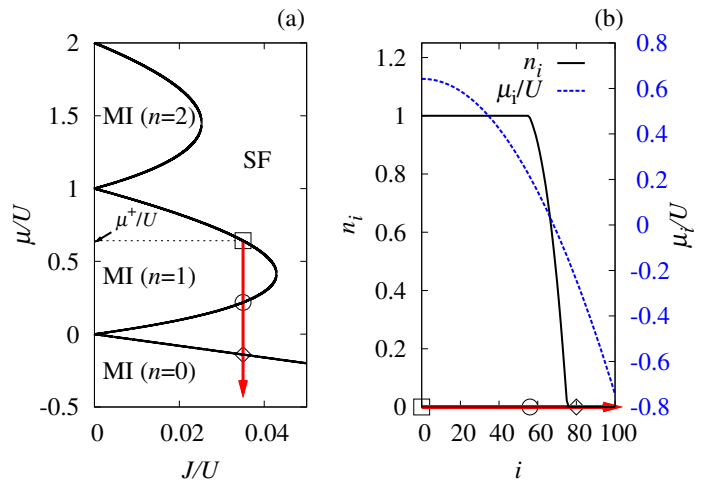}
\caption{(Color online) (a) Mean-field phase diagram of the Bose-Hubbard 
model in 2D. The red arrow follows the different phases occurring in the trapped 
system as one moves away from the center of the trap. (b) The density profile 
as obtained from the LDA with the largest Mott domain plotted together with the spatially 
varying chemical potential $\mu_i/U$.}
\label{fig:LDA}
\end{figure}

We are interested in the expansion of Mott-insulating states, though we know {\it a priori}
 that, in presence of a trapping potential, such Mott states are inevitably
surrounded by superfluid regions. Nevertheless, one can aim at finding a setup 
for which those Mott domains are as large as possible for a given value of $J/U$. Our
approach to achieving large Mott domains is based on the local-density approximation 
(LDA). The LDA assumes that the local properties of the confined system
can be mapped to those properties of the homogeneous system with the same value of 
$J/U$ and an effective chemical potential corresponding to the spatially varying 
 $\mu_i=\mu-V_i$ at site $i$ in the trap. One can, therefore,  
get an approximate picture of what the trapped system looks like based on the phase 
diagram of the homogeneous system.
Within the LDA, in order to get the largest Mott domain, it is enough to set the 
chemical potential at the center of the trap  to be the largest chemical
potential of the homogeneous system that allows for a Mott insulator with density $n=1$.
That is just the value of the chemical potential $\mu^{+}$ right at the transition 
from the Mott insulator with density $n=1$ to the superfluid with $n>1$.
This is based on the observation that, as the distance from the center of the trap is 
increased, the effective chemical potential of the homogeneous system decreases. 
In 2D, the critical value  $\mu^{+}$ can be determined within 
the mean-field theory by finding the largest solution for $\mu$  from the following 
relation that determines the boundaries of the Mott lobes \cite{sachdev_00},
\begin{equation}\label{eq:boundary}
\dfrac{2\left(J_x+J_y \right)}{U}= \dfrac{\left(n-\mu/U \right)\left( \mu/U-n+1 \right) }{1+\mu/U},
\end{equation}
where $J_x$ and $J_y$ are the hopping parameters along the $x$ and $y$ directions, 
respectively. These ideas are illustrated in Fig.~\ref{fig:LDA}, where the mean-field 
phase diagram of the homogeneous 2D system is depicted [Fig.~\ref{fig:LDA}(a)], as well 
as the density profile across the center of a trapped system that contains the largest 
Mott domain as obtained from the LDA [Fig.~\ref{fig:LDA}(b)].

We should stress that, once $\mu^{+}$ has been identified, we select a value of $\mu$
close to $\mu^{+}$ and solve Eq.~\eqref{mfr0} and/or Eq.~\eqref{meanfbose1} to obtain 
the initial state. Hence our initial state is always the exact ground state within the 
mean-field approximation, i.e., it is not the LDA ground state. This is the way the 
initial wave function in Eq.~\eqref{productState} is constructed for all our expansions. 

\section{Expansion in one direction} \label{laser}

Let us first consider the expansion of a Mott insulator along one direction in the lattice. 
We study the case in which the initial state is confined only along the $x$ direction,
i.e., $V_y=0$, while $V_x$ is kept finite, but the system has open boundary conditions 
in the $y$ direction (a box trap). We allow for the hopping amplitudes to be anisotropic, 
namely, that the hopping along the $x$ direction ($J_x$) differs from the one along the 
$y$ direction ($J_y$). We then study the expansion for several values of $U/J_x$, large enough 
so that the Mott-insulating phase exists, as well as different values of $\eta=J_y/J_x$. In all cases 
considered in this section, the expansion takes place in a lattice with $200\times 40$ sites.
As stated in the Introduction, this study extends the results presented in Ref.~\cite{hen_rigol_10} 
on the expansion of hard-core bosons in three-dimensional lattices to the expansion of 
soft-core bosons confined in two-dimensional lattices. The initial state is selected such that 
the Mott-insulating domain is the largest possible for given values of $U/J_x$, 
$\eta$, and $V_x/J_x$, as explained in the previous section. 

\begin{figure}[!t]
\includegraphics*[width=0.48\textwidth]{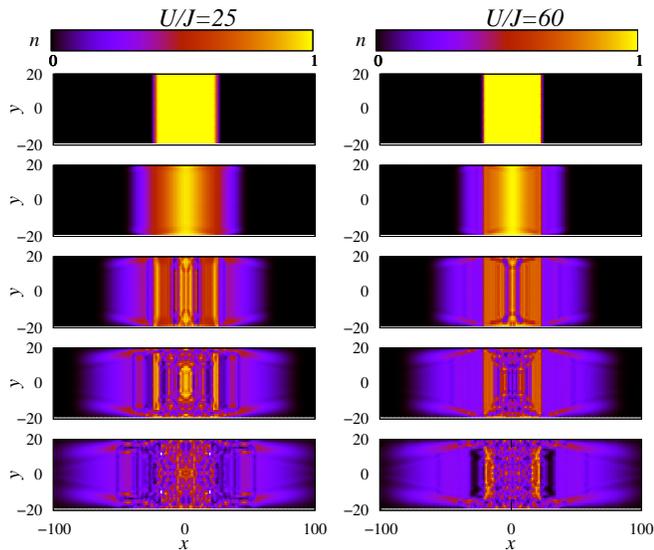}
\caption{(Color online) Comparison between the time evolution of the density profile
across the trap of systems with $U/J_x=25$ (left) and $U/J_x=60$ (right), at
$t=0,\ 10,\ 20, \ 30$ and $40$ (from top to bottom) following the release from
traps with $V_x/J_x=0.03$ and $V_x/J_x=0.08$, respectively. $\eta=0.3$ and  the time is reported in
units of $\hbar/J_x$.}
\label{fig:anisoDENS}
\end{figure}

In Fig.~\ref{fig:anisoDENS}, we show the time evolution of the density profiles of systems 
with $U/J_x=25$ and $60$, for $\eta=0.3$. In the initial state ($t=0$), both systems exhibit 
large Mott-insulating domains in the center of the lattice ($x=0$), surrounded by small 
superfluid regions, the latter being larger for the smallest interaction strength, 
i.e., $U/J_x=25$. For $t>0$, the particles are allowed to expand 
by switching off the confining potential along the $x$ direction. As a result, the 
Mott-insulating domains melt. This process is seen to be qualitatively similar for 
the two values of $U/J_x$ considered in Fig.~\ref{fig:anisoDENS}. Note that during the time evolution, due 
to the symmetry of the initial state, the system expands only along the $x$ direction, and the 
density profile maintains the reflection symmetry with respect to both $x$ and $y$ axes.

\begin{figure}[!t]
\includegraphics*[width=0.48\textwidth]{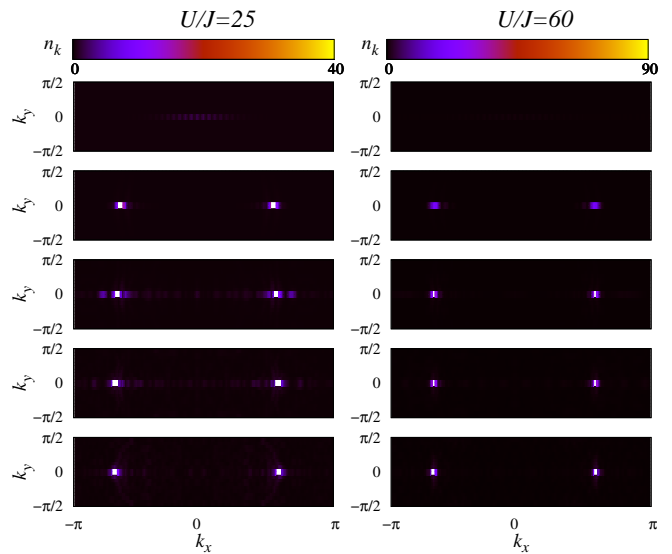}
\caption{(Color online) Comparison between the time evolution of the momentum distribution
function of systems with $U/J_x=25$ (left) and $U/J_x=60$ (right), at $t=0,\ 10,\ 20, \ 30$ and $40$
(from top to bottom) following the release from traps with $V_x/J_x=0.03$ and $V_x/J_x=0.08$,
respectively. $\eta=0.3$ and the time is reported in units of $\hbar/J_x$.}
\label{fig:anisoMOM}
\end{figure}

The corresponding momentum distributions $n_{\mathbf{k}}$ of the systems we have just described 
are presented in Fig.~\ref{fig:anisoMOM}. At $t=0$, although barely 
observable in the plots (first row in Fig.~\ref{fig:anisoMOM}), there is a sizable population of 
low momentum modes, because of the presence of the inevitable superfluid regions 
surrounding the initial Mott insulators. Those modes are embedded in a roughly homogeneous 
background of low-populated momenta. The latter are the result of the large
Mott-insulating domains present in the initial states (top panels in Fig.~\ref{fig:anisoDENS}). 
Remarkably, shortly after the expansion has started, particles get 
redistributed in momentum space, and they tend to fill up states surrounding two 
well-defined momentum values $\mathbf{k}=\left(\pm  k_x^* , 0\right)$. This effect, 
which can be clearly observed in all panels with $t>0$ in Fig.~\ref{fig:anisoMOM}, is already 
present at very early times $t<10$. As the expansion goes on, the values of the momentum of the 
highly populated modes become time independent. Their finite momenta reflects the motion of the 
two groups of particles moving away from the center of the systems in Fig.~\ref{fig:anisoDENS}. 
We find that the structure around the momentum peaks, better seen in the left column in 
Fig.~\ref{fig:anisoMOM} around $\mathbf{k}$, disappears as the value of $U/J_x$ is increased, 
and only sharp peaks are left. 
This is apparent by comparing the momentum distribution functions between $t=20$ and $t=40$ in 
Fig.~\ref{fig:anisoDENS}. There, the peaks for $U/J_x=60$ are much sharper than those for $U/J_x=25$. 

To a very good approximation, on can predict the values of $\mathbf{k}$ seen in Fig.~\ref{fig:anisoMOM} 
using energy and momentum conservation arguments. In order to do that, we first realize that, if $U/J_x$ 
is large enough, double or higher occupancies of the lattice sites become energetically too costly and 
can be, to a first approximation, ignored. The system can then be thought as composed of hard-core 
bosons. Following Ref.~\cite{hen_rigol_10}, the kinetic energy can then be written as
\begin{eqnarray}
 E_\textrm{kin}=\sum_{\mathbf{k} }\ \epsilon_{\mathbf{k}} n_{ \mathbf{k}},
\label{KE1aniso}
\end{eqnarray}
where $n_{\mathbf{k}}$ is the momentum distribution function at any given time, 
$\mathbf{k}=(k_x,k_y)$, and $\epsilon_{\mathbf{k}}$ is the dispersion relation 
\begin{eqnarray}
 \epsilon_{\mathbf{k}} =-2J_x\cos{k_x} -2J_y \cos{k_y}.
\label{eq:disprelaniso}
\end{eqnarray}
In order to further simplify the calculations, one can assume that the system before 
the expansion is a pure Mott insulator, thus neglecting the small superfluid regions that are always
present in the trap, and implying that the initial kinetic energy is $E_\textrm{kin}\approx0$. 
(Note that if double occupancies are neglected, the kinetic energy is essentially the total energy 
of the system after the trap has been turned off.) We further recall the observation that, in 
Fig.~\ref{fig:anisoMOM}, many particles tend to populate modes with 
$\mathbf{k}=\left(\pm  k_x^* , 0\right)$, from which we understand that the $y$ component 
of $\mathbf{k}$ is zero due to the symmetry of the expansion, 
which occurs only along the $x$ direction. We can now ask, within our simplified picture
of the systems presented in Figs.~\ref{fig:anisoDENS} and \ref{fig:anisoMOM}, which modes  
[with ($k^*_x,0$)] could be populated by all the particles after the expansion while still  
conserving the energy. (We are assuming that interactions during the 
expansion redistribute the energy in such a way that all particles condense into 
a single mode.) From Eq.~\eqref{KE1aniso}, it follows that
\begin{eqnarray}
 \cos{k^*_x} =-\eta. 
\label{eq:momcond}
\end{eqnarray}
If, in addition to energy conservation, we enforce conservation of the total lattice 
momentum, or the symmetry of the expansion for that matter, we conclude that 
$\mathbf{k}=\left(\pm  k_x^* , 0\right)$, where $k^*_x=\arccos(-\eta)$.

\begin{figure}[!t]
\centering
\includegraphics[width=0.48\textwidth]{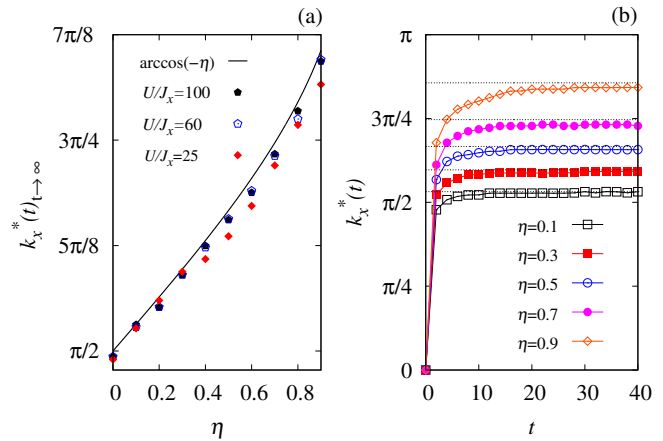}
\caption{(Color online) (a) Location of the momentum peak $k^*_x$ after a long expansion time 
obtained from the mean-field calculation as a  function of the hopping ratio $\eta$, for $U/J_x=25, \ 60,$ 
and $100$ (points). The prediction of Eq.~\eqref{eq:momcond} is depicted as a solid line. 
(b) The time evolution of the location of the momentum peak for $U/J_x=60$ and several values of 
$\eta$. The horizontal dashed-dotted lines show the values predicted in Eq.~\eqref{eq:momcond}. The time 
is reported in units of $\hbar/J_x$.} 
\label{fig:peak}
\end{figure}

From Eq.~\eqref{eq:momcond}, it follows that, in the limit of $U/J_x\gg1$, the 
location of the $\mathbf{k}$ modes that are populated during the expansion 
depends only on the ratio $\eta$. Remarkably, we find that the peaks 
in Fig.~\ref{fig:anisoMOM} agree very well with that prediction [Eq.~\eqref{eq:momcond}]. 
This occurs despite the fact that, in Fig.~\ref{fig:anisoMOM}, the results were obtained 
for finite values of $U/J_x$ and in the presence of small superfluid domains around the
Mott insulator. In Fig.~\ref{fig:peak}(a), we report a complete study of the dependence of the 
location of the peaks in the mean-field calculations as a function of $\eta$, for three 
values of $U/J_x$ (points), together with the predictions of Eq.~\eqref{eq:momcond} (solid line). 
For the smallest values of $U/J_x$ that allow for the realization of a Mott insulator, 
one expects some dependence of the location of the peaks on $U/J_x$, because the dispersion 
relation is expected to exhibit the largest differences from Eq.~\eqref{eq:disprelaniso}. 
The deviations of the mean-field results from Eq.~\eqref{eq:momcond} are, however, very 
small for small values $\eta$. The largest deviations, for all values of $U/J_x$, are observed 
for the largest values of $\eta$. This occurs because as $\eta$ is increased, the superfluid 
wings surrounding the Mott insulator increase their size; as a result, the total kinetic energy 
is lower than that of the Mott insulator, and one of our assumptions ($E_\textrm{kin}\approx0$) 
starts to fail. Nevertheless, as Fig.~\ref{fig:peak}(a) shows, for the largest values 
of $\eta$, the differences are still relatively small even for the smallest values of $U/J_x$ that 
support a Mott insulator. They also can be seen to decrease when increasing $U/J_x$. For 
$U/J_x=100$, the results are barely distinguishable from the analytic results for the hard-core 
limit. Overall, the fact that only for large values of $U/J_x$ one can realize a Mott insulator
in the initial state allows for the hard-core description to be a good approximation in all 
cases analyzed here.
 
We conclude this section by showing, in Fig.~\ref{fig:peak}(b), results for the location of the 
momentum peak as function of time, for $U/J_x=60$. That figure makes clear that, as time passes, the 
location of the peak rapidly approaches the values predicted by Eq.~\eqref{eq:momcond} (dotted horizontal 
lines). It also shows that it takes a shorter time to reach those values for smaller values of $\eta$. 
In our calculations, the value of $k^*_x$ is determined as the local center of mass of the momentum 
distribution function around the position of the brightest peak.

\section{Expansion in all directions} \label{isotropic}

We now examine the situation in which the initial Mott insulator is allowed to expand 
in all directions. For such a case, we consider $V_x=V_y=V$ and $J_x=J_y=J$. We turn off 
the trap at $t=0$, and study the time evolution of the density and 
momentum distribution for systems with $U/J=25,\ 30,\ 40,\ 50,\ 
60,\ 80,$ and $100$. Note that the Mott insulator can only be achieved in systems 
with large enough values of $U/J$ ($U/J\gtrsim 23.2$ within the mean-field approach 
in 2D).

\begin{figure}[!t]
\includegraphics*[width=0.35\textwidth]{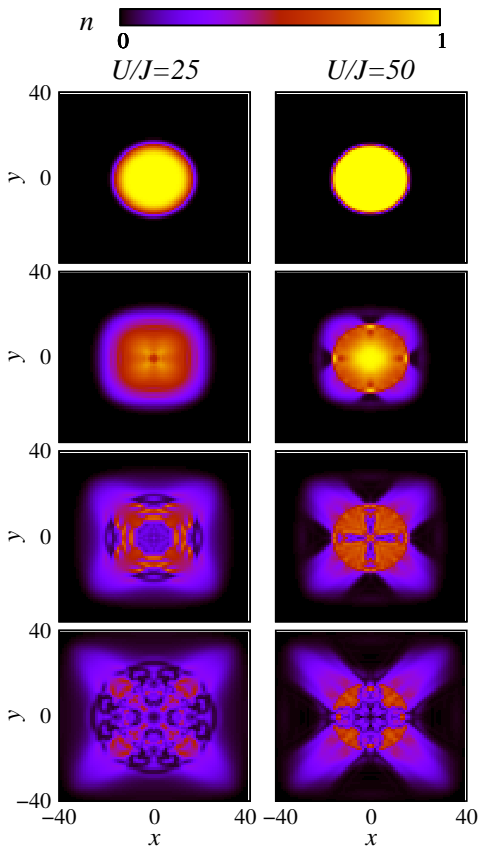}
\caption{(Color online) Comparison between the time evolution of the density profile 
across the trap of systems with $U/J=25$ (left) and $U/J=50$ (right), at 
$t=0,\ 4,\ 8,$ and $12$ (from top to bottom) following the release from 
traps with $V/J=0.053$ and $V/J=0.161$, respectively. The time is reported in 
units of $\hbar/J$.}
\label{fig:isoDENS}
\end{figure}

Figure \ref{fig:isoDENS} depicts the time evolution of the density profile of the
expanding bosons with $U/J=25$ (left) and $U/J=50$ (right). Interestingly, although 
the rightmost case has an interaction energy that is twice that of the leftmost one, 
the expansion is somehow similar in both cases. At the moment the trap is turned off 
($t=0$), the Mott-insulator domain is easily discernible from the superfluid one 
surrounding it. The lattice sites in the vicinity of the center of the system
 have unit occupancies and are surrounded by a thin ring with density smaller 
than one. Note that, as expected from the way we construct the initial state 
(see Sec.~\ref{sec:inistate}), the superfluid ring in the system with $U/J=50$ is 
smaller than the one in the system with $U/J=25$.

After turning off the trap, the bosonic cloud expands and the Mott insulator melts. 
In Fig.~\ref{fig:isoDENS}, it is remarkable that the expansion 
along the diagonals is always faster than that along the $x$ and $y$ axes. This leads
to a square-like density profile for $t>0$. Such a behavior is accentuated as $U/J$ 
increases, as seen in the right panels in Fig.~\ref{fig:isoDENS}, where the expansion 
occurs almost entirely along the diagonals. 

\begin{figure}[!t]
\includegraphics*[width=0.35\textwidth]{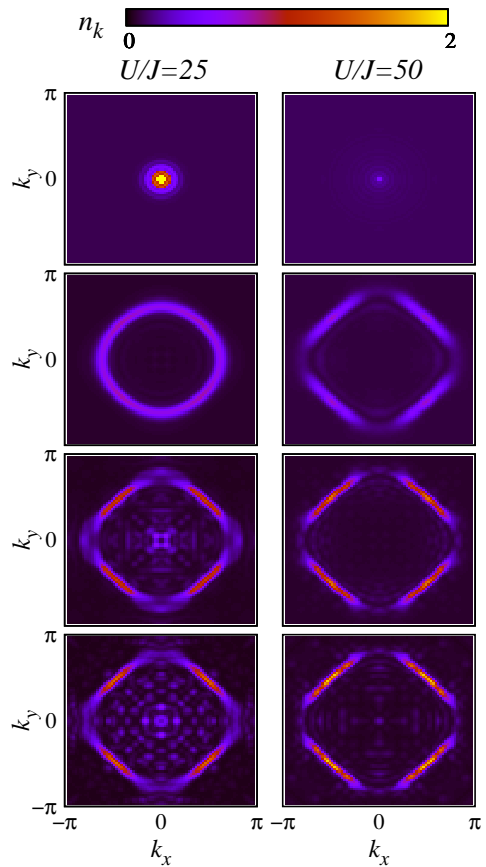}
\caption{(Color online) Comparison between the time evolution of the momentum distribution 
function of systems with $U/J=25$ (left) and $U/J=50$ (right), at $t=0,\ 4,\ 8,$ and $12$ 
(from top to bottom) following the release from traps with $V/J=0.053$ and $V/J=0.161$, 
respectively. The time is reported in units of $\hbar/J$.}
\label{fig:isoMOM}
\end{figure}

\begin{figure*}[!t]
\includegraphics*[width=0.80\textwidth]{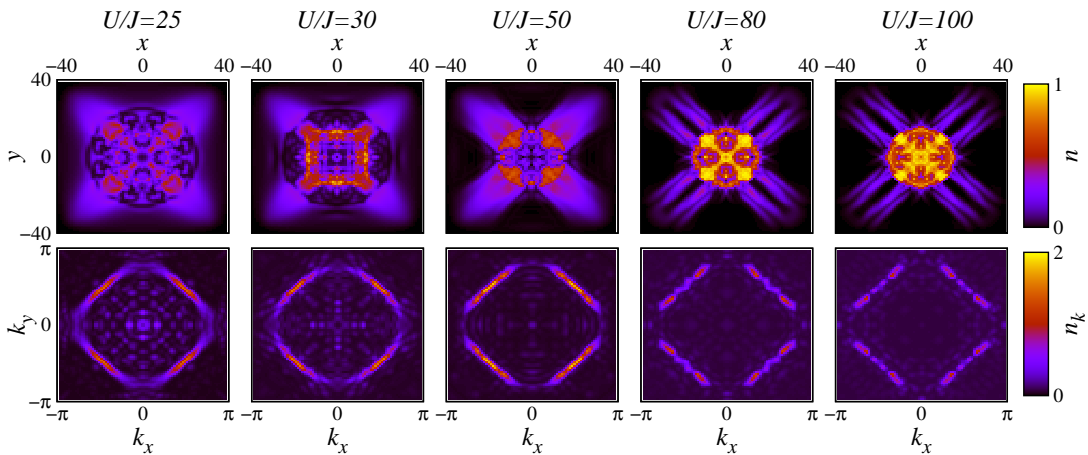}
\caption{(Color online) Top row: Comparison between the density profiles
of systems with $U/J=25,\ 30,\ 50,\ 80,$ and $100$ (from left to right), at $t=12$ 
following the release from traps with $V/J=0.053,\ 0.0764,\ 0.161,\ 0.278$ and $0.356$, 
respectively. Bottom row: Corresponding momentum distribution functions. The time is reported in units of $\hbar/J$.}
\label{fig:isoALL}
\end{figure*}

The evolution of the momentum distribution function, corresponding to the density profiles 
depicted in Fig.~\ref{fig:isoDENS}, is shown in Fig.~\ref{fig:isoMOM}. At $t=0$, peaks 
can be seen in the center of the momentum distribution function. They correspond to the 
bosons that are in the superfluid rings surrounding the Mott insulator. The difference 
between the size of those rings for $U/J=25$ (Fig.~\ref{fig:isoDENS}, left) and $U/J=50$ 
(Fig.~\ref{fig:isoDENS}, right) is apparent in the initial momentum distribution, 
as the height of the peak for the $U/J=25$ case is significantly larger than that 
in the $U/J=50$ case. Surprisingly, for $t>0$, these peaks rapidly evolve into circular-like 
structures, which in turn, become square-like as time passes (Fig.~\ref{fig:isoMOM}). 
The initial ring structure is less evident in the $U/J=50$ case at $t=4$ because the 
formation of the square-like structure occurs faster as the value of $U/J$ is increased.
Also, the structure discernible in the momentum distribution function for small values of 
$k_x$ and $k_y$ (bottom panels for $U/J=25$) disappears as the value of $U/J$ is 
increased (see how much weaker it is in bottom panels for $U/J=50$) and most of 
the bosons have momenta that lay within the edges of the square. The existence of this 
structure, in which bosons remain with small values of $k_x$ and $k_y$, can be directly 
related to the size of the initial superfluid domain, which, as said before, decreases 
as the system approaches the hard-core limit.

A comparison between the density and momentum distributions of systems with different values
of $U/J$ ($U/J=25,\ 30,\ 50,\ 80,$ and $100$), at the time the bosons reach the boundaries 
of our 80 $\times$ 80 lattice ($t=12$), is presented in  Fig.~\ref{fig:isoALL}. These plots make 
evident that the expansion mainly occurs along the diagonals as $U/J$ increases and, at the 
same time, the momentum distribution function acquires a clearer square shape.

\begin{figure}[!h]
\includegraphics*[width=0.42\textwidth]{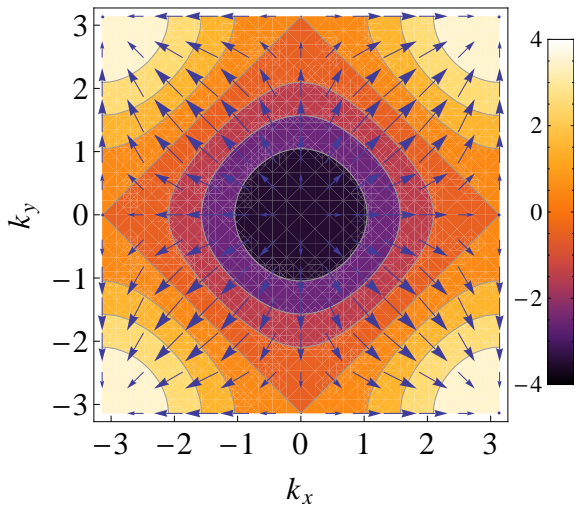}
\caption{(Color online) Depiction of the group velocity vector field, given by Eq.~\eqref{velocity} 
in $k$-space, along with a series of contour plots satisfying the equation $-2(\cos{k_x}+\cos{k_y})=w/J$, where $w/J=-2,\ -1.5,\ -1,\ -0.5,\ 0,\ 0.5,\ 1,$ and $2$.}
\label{fig:isoCOND}
\end{figure}

An understanding of the behavior exhibited by these systems, when allowed to expand in all
directions, can be gained following a similar line of reasoning as done in the previous 
section. If $U/J\gg1$, a good description is provided by the hard-core boson picture. For 
hard-core bosons, after the trapping potential is turned off, 
the total energy is the kinetic energy, just as in the expansion in one direction.  
The dispersion relation $\epsilon_{\mathbf{k}}$ in the isotropic case reduces to  
\begin{eqnarray}
 \epsilon_{\mathbf{k}} =-2J(\cos{k_x} +\cos{k_y}).
\label{eq:disprel}
\end{eqnarray}
We now assume, once again to simplify the argument, that the system before the expansion is 
a perfect Mott insulator, i.e., all the particles are localized 
in their respective sites, and occupy all possible momenta in  $\mathbf{k}$
space. Therefore, the kinetic energy is again
$E_\textrm{kin}=0$, and it is conserved during the expansion. Furthermore, 
let us assume that, after long time, the majority of bosons will occupy 
(``condense'' to) modes with particular $\mathbf{k}$ vectors (as seen in the mean-field 
calculations). Energy conservation is guaranteed by the assumption that the 
energy of each occupied mode vanishes. Under those assumptions, 
the $\mathbf{k}$ vector of the modes populated by the bosons can  
be calculated from the dispersion relation as
\begin{eqnarray}
\epsilon_{\mathbf{k}}=-2J (\cos{k_x}+\cos{k_y})=0.
\label{condition}
\end{eqnarray}
The solution to this equation is given by the square-like contour plot seen in
Fig.~\ref{fig:isoCOND}, for the case with $w=0$. Analytically, 
such a contour plot is described by the expressions
\begin{align}
\qquad k_y=\pi \pm k_x,
\qquad k_y=-\pi \pm k_x.
\label{lines}
\end{align}

Remarkably, the momentum distribution of all systems in Fig.~\ref{fig:isoALL} 
has converged to a distinctive square-like structure. It closely resembles
the prediction of Eq.~\eqref{lines}, as represented in Fig.~\ref{fig:isoCOND}. The deviations 
from the exact results in Eq.~\eqref{lines} can be understood to be the result of (i) the 
finite values of $U/J$ in all our calculations, and (ii) the fact that the Mott-insulating 
domains with $n=1$ are always surrounded by superfluid ones. Once again, the first modifies 
the dispersion relation in Eq.~\eqref{eq:disprel}, and the second one reduces the energy from 
the value assumed in Eq.~\eqref{condition}. As clearly seen in Fig.~\ref{fig:isoALL}, when 
$U/J$ increases and the superfluid domains in the initial state reduce their size, the 
momentum distribution becomes closer to the square-like contour defined in Eq.~\eqref{lines}.
The robustness of the results, despite points (i) and (ii) above, is remarkable. 
Hence we expect that they should be reproducible in experiments with ultracold bosons
even at finite, but low, temperatures.

The analysis above also allows one to understand the expansion pattern seen in the density profiles
in Figs.~\ref{fig:isoDENS} and \ref{fig:isoALL}. The density profiles for $t>0$ are determined by
the velocities with which bosons expand across the lattice. In the hard-core limit, the group velocity is
\begin{align}
\bold{v}_g=\bold{\nabla} \epsilon_{\mathbf{k}}  =2J (\sin{k_x}\hat{x}+\sin{k_y}\hat{y}),
\label{velocity}
\end{align}
which produces the vector field plot seen in Fig.~\ref{fig:isoCOND}. It is clear that the group 
velocities, which are perpendicular to the contour plots, increase in magnitude for those 
$k$ modes along the diagonal. (The maximum group velocity occurs when 
$\epsilon_{\mathbf{k}}=0$ along the diagonal.) Therefore, if such modes are highly populated, 
the net effect is that the system will expand preferably along the directions indicated by 
those group velocity vectors. Indeed, the high population of such modes is apparent in the momentum 
distribution functions depicted in Fig.~\ref{fig:isoALL}. The low population of the modes in the 
vertices of the square indicates the suppression of particles with momenta close to 
$(\pm \pi,0)$ and $(0,\pm \pi)$, which have vanishing group velocities. In Fig.~\ref{fig:isoALL}, 
the accuracy of this description can be seen to improve with increasing $U/J$. It is 
interesting to note that the system expands at the greatest speed possible by populating 
the modes with the highest group velocity that are consistent with the conservation of the energy, as
can be concluded from the momentum distributions and comparing them with the analytical results.

We have already shown that, during the expansion from a mostly Mott-insulating state, bosons 
tend to populate momentum modes around well-defined values. Results have been presented for 
mean-field calculations and supported by analytic arguments. One can now wonder whether all 
these highly populated modes are coherently coupled forming a simple Bose-Einstein condensate, 
or whether they are part of fragmented condensates, of quasicondensates, or their occupation 
is simply $O(N_b^0)$ but their number increases with increasing system size. In strongly correlated 
systems, it is possible that modes with different momenta couple to each other and still form a 
simple Bose-Einstein condensate.

In the presence of interactions, condensation can be understood as effective single-particle
states (natural orbitals) being macroscopically populated. Following Penrose and Onsager \cite{penrose56}, 
one needs to diagonalize the one-particle density matrix $\rho_{i,j}$ and study its eigenvalues 
$\lambda_{\eta}$
\begin{equation}
 \sum_{j} \rho_{i,j} \varphi^{\eta}_j=\lambda_{\eta} \varphi^{\eta}_i.
\end{equation}
(i) If only one of them scales proportionally to the total number of bosons in the system ($N_b$), 
one has simple Bose-Einstein condensation. (ii) If more than one of them scales proportionally to $N_b$, 
this means that fragmentation occurs \cite{leggett01}. (iii) One can also have many of them
that scale proportionally to $N_b^\alpha$ ($0<\alpha<1$), and one would then say that 
quasicondensation occurs. (iv) Finally, if all of them scale as $O(N_b^0)$, the system does not 
exhibit Bose-Einstein condensation or quasicondensation. Interestingly, exact results for 
expanding hard-core bosons in one dimension \cite{rigol_muramatsu_05eHCBb} have already shown 
that the largest eigenvalue of the one-particle density matrix in a system out of equilibrium 
can diverge ($\propto \sqrt{N_b}$ in that case), while still being composed by bosons with many 
different momenta. In those systems, there were as many populated momenta as those available 
in the Fermi sea of the noninteracting fermions to which hard-core bosons were mapped.

Within the mean-field approximation, the elements of the one-particle density 
matrix reduce to 
\begin{align}
\rho_{i,j}=\Phi_i^*\Phi_j + \delta_{i,j}\left( n_i-\left|\Phi_i\right|^2 \right).
\label{eq:dm}
\end{align}
The expression in Eq.~\eqref{eq:dm}, which holds because of the intrinsic form of the Gutzwiller 
state in Eq.~\eqref{productState}, implies that the one-particle density matrix can only attain 
at most one eigenvalue which grows as the total number of particles grows; thus if fragmentation 
or quasicondensation occurs in a real system, the Gutzwiller mean-field approach will fail to 
reproduce them \cite{buchhold_11}. A scenario in which all $\lambda$'s are $O(N_b^0)$, on the other 
hand, can of course be reproduced by this approximation (e.g., the Mott insulator at zero temperature 
and the normal phase at high temperature).

\begin{figure}[h]
\centering
\includegraphics*[width=0.35\textwidth]{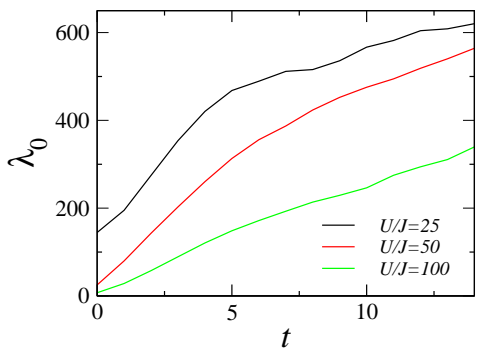}
\caption{(Color online) Comparison between the time evolution of the occupation of the highest 
populated natural orbital $\lambda_0$ for systems with $U/J=25$ (black), $U/J=50$ (red), and 
$U/J=100$ (green), after turning off traps with $V/J=0.053,\ 0.162,$ and $0.356$, respectively. The time is reported in units of $\hbar/J$.}
\label{fig:natorbit}
\end{figure}

In Fig.~\ref{fig:natorbit}, we show the time evolution of the largest eigenvalue $\lambda_0$ 
of $\rho_{i,j}$ for three different values of $U/J$ ($U/J=25,\ 50,$ and $100$) in our systems.
Due to the presence of the superfluid rings in the initial state, $\lambda_0$ is not always small
at $t=0$. (As expected, its value at $t=0$ decreases as $U/J$ is increased.) For all the systems
studied in this work, we find that the occupation of the highest populated natural orbital always grows 
during the expansion (Fig.~\ref{fig:natorbit}), while the occupation of the other natural orbitals 
remains negligible at all times. In addition, we have verified that $\lambda_0$ always increases 
proportionally to the total number of particles. Hence, at least within the mean-field approximation, 
we find that one natural orbital is macroscopically populated. This means that one of the scenarios 
(i)-(iii) could be relevant to the physics of the real systems. Experiments and/or other theoretical 
approaches will need to be used to identify which of those scenarios (if any) is the correct one.

\section{Conclusions} \label{conclusions}
We have presented a detailed study of the expansion of $n=1$ bosonic Mott insulators in
two-dimensional optical lattices. In all cases studied here, the initial states were prepared 
in such a way that the largest fraction possible of the bosons would be part of the Mott-insulating 
domain. Two setups were considered: the expansion along one direction of an anisotropic 
system, which was initially trapped only along the $x$ direction, and the expansion of an isotropic 
system, which was allowed to expand in all directions. In the first case, we have shown that the 
expansion of the Mott insulator leads to a high population of two well-defined momentum states. 
Those momenta were found to be controlled by the ratio of the hopping amplitudes along the $x$ and 
$y$ directions.  In the second case, the expansion of a Mott insulator in all directions  
produced a simple condensate composed by bosons with many different momenta. Given the approximated nature 
of our calculations, experiments, and/or unbiased theoretical work, will need to verify our findings 
and address the nature of the condensation seen within the Gutzwiller mean-field approximation. 
For the two set ups mentioned before, we also presented analytical arguments to determine which 
momentum modes are populated during the expansion. 

It is important to notice that the phenomena we have observed here arise because of a subtle 
interplay between quantum tunneling and strong correlations. Both effects play a crucial role in the 
generation of the condensates as, on the one hand, tunneling is necessary for the system
to expand, and on the other hand, interactions are necessary for the bosons to redistribute 
the energy and couple coherently. In the absence of interactions, the system still expands, but
the momentum distribution remains unchanged from its initial value, i.e., it 
remains featureless if one starts from purely localized particles in the lattice sites
(a Fock state), and no condensation occurs.

Finally, our study supports the conclusion in Ref.~\cite{hen_rigol_10}, and generalizes it to 
the soft-core boson case, that strongly correlated atom lasers can be generated from the 
expansion of Mott-insulating states. Furthermore, after the results in 
Sec.~\ref{laser}, one can see that the momenta of those matter waves can be controlled by 
changing the ratio between the hopping parameters in the different directions in the lattice. 
In addition, if one cannot change those, in an isotropic lattice (Sec.~\ref{isotropic}), 
the momenta of the lasers can be controlled by changing the direction along which the Mott 
insulator is allowed to expand. In the latter setup, the best results would be obtained by 
constraining the Mott insulator to expand along the diagonals in the lattice.

\begin{acknowledgments}
This work was supported by the U.S. Office of Naval Research.
\end{acknowledgments}


\begin{thebibliography}{44}
\expandafter\ifx\csname natexlab\endcsname\relax\def\natexlab#1{#1}\fi
\expandafter\ifx\csname bibnamefont\endcsname\relax
  \def\bibnamefont#1{#1}\fi
\expandafter\ifx\csname bibfnamefont\endcsname\relax
  \def\bibfnamefont#1{#1}\fi
\expandafter\ifx\csname citenamefont\endcsname\relax
  \def\citenamefont#1{#1}\fi
\expandafter\ifx\csname url\endcsname\relax
  \def\url#1{\texttt{#1}}\fi
\expandafter\ifx\csname urlprefix\endcsname\relax\def\urlprefix{URL }\fi
\providecommand{\bibinfo}[2]{#2}
\providecommand{\eprint}[2][]{\url{#2}}

\bibitem[{\citenamefont{Bloch et~al.}(2008)\citenamefont{Bloch, Dalibard, and
  Zwerger}}]{bloch_dalibard_review_08}
\bibinfo{author}{\bibfnamefont{I.}~\bibnamefont{Bloch}},
  \bibinfo{author}{\bibfnamefont{J.}~\bibnamefont{Dalibard}}, \bibnamefont{and}
  \bibinfo{author}{\bibfnamefont{W.}~\bibnamefont{Zwerger}},
  \bibinfo{journal}{Rev. Mod. Phys.} \textbf{\bibinfo{volume}{80}},
  \bibinfo{pages}{885} (\bibinfo{year}{2008}).

\bibitem[{\citenamefont{Cazalilla et~al.}()\citenamefont{Cazalilla, Citro,
  Giamarchi, Orignac, and Rigol}}]{cazalilla_citro_11}
\bibinfo{author}{\bibfnamefont{M.~A.} \bibnamefont{Cazalilla}},
  \bibinfo{author}{\bibfnamefont{R.}~\bibnamefont{Citro}},
  \bibinfo{author}{\bibfnamefont{T.}~\bibnamefont{Giamarchi}},
  \bibinfo{author}{\bibfnamefont{E.}~\bibnamefont{Orignac}}, \bibnamefont{and}
  \bibinfo{author}{\bibfnamefont{M.}~\bibnamefont{Rigol}},
  \bibinfo{note}{arXiv:1101.5337 (2011), To be published in Rev. Mod. Phys.}.

\bibitem[{\citenamefont{Greiner
  et~al.}(2002{\natexlab{a}})\citenamefont{Greiner, Mandel, Esslinger,
  H\"ansch, and Bloch}}]{greiner_mandel_02a}
\bibinfo{author}{\bibfnamefont{M.}~\bibnamefont{Greiner}},
  \bibinfo{author}{\bibfnamefont{O.}~\bibnamefont{Mandel}},
  \bibinfo{author}{\bibfnamefont{T.}~\bibnamefont{Esslinger}},
  \bibinfo{author}{\bibfnamefont{T.~W.} \bibnamefont{H\"ansch}},
  \bibnamefont{and} \bibinfo{author}{\bibfnamefont{I.}~\bibnamefont{Bloch}},
  \bibinfo{journal}{Nature} \textbf{\bibinfo{volume}{415}}, \bibinfo{pages}{39}
  (\bibinfo{year}{2002}{\natexlab{a}}).

\bibitem[{\citenamefont{Spielman et~al.}(2007)\citenamefont{Spielman, Phillips,
  and Porto}}]{spielman_phillips_07}
\bibinfo{author}{\bibfnamefont{I.~B.} \bibnamefont{Spielman}},
  \bibinfo{author}{\bibfnamefont{W.~D.} \bibnamefont{Phillips}},
  \bibnamefont{and} \bibinfo{author}{\bibfnamefont{J.~V.} \bibnamefont{Porto}},
  \bibinfo{journal}{Phys. Rev. Lett.} \textbf{\bibinfo{volume}{98}},
  \bibinfo{pages}{080404} (\bibinfo{year}{2007}).

\bibitem[{\citenamefont{St\"oferle et~al.}(2004)\citenamefont{St\"oferle,
  Moritz, Schori, K\"ohl, and Esslinger}}]{stoferle_moritz_04}
\bibinfo{author}{\bibfnamefont{T.}~\bibnamefont{St\"oferle}},
  \bibinfo{author}{\bibfnamefont{H.}~\bibnamefont{Moritz}},
  \bibinfo{author}{\bibfnamefont{C.}~\bibnamefont{Schori}},
  \bibinfo{author}{\bibfnamefont{M.}~\bibnamefont{K\"ohl}}, \bibnamefont{and}
  \bibinfo{author}{\bibfnamefont{T.}~\bibnamefont{Esslinger}},
  \bibinfo{journal}{Phys. Rev. Lett.} \textbf{\bibinfo{volume}{92}},
  \bibinfo{pages}{130403} (\bibinfo{year}{2004}).

\bibitem[{\citenamefont{Greiner
  et~al.}(2002{\natexlab{b}})\citenamefont{Greiner, Mandel, H\"ansch, and
  Bloch}}]{greiner_mandel_02b}
\bibinfo{author}{\bibfnamefont{M.}~\bibnamefont{Greiner}},
  \bibinfo{author}{\bibfnamefont{O.}~\bibnamefont{Mandel}},
  \bibinfo{author}{\bibfnamefont{T.~W.} \bibnamefont{H\"ansch}},
  \bibnamefont{and} \bibinfo{author}{\bibfnamefont{I.}~\bibnamefont{Bloch}},
  \bibinfo{journal}{Nature} \textbf{\bibinfo{volume}{419}}, \bibinfo{pages}{51}
  (\bibinfo{year}{2002}{\natexlab{b}}).

\bibitem[{\citenamefont{Kinoshita et~al.}(2006)\citenamefont{Kinoshita, Wenger,
  and Weiss}}]{kinoshita_wenger_06}
\bibinfo{author}{\bibfnamefont{T.}~\bibnamefont{Kinoshita}},
  \bibinfo{author}{\bibfnamefont{T.}~\bibnamefont{Wenger}}, \bibnamefont{and}
  \bibinfo{author}{\bibfnamefont{D.~S.} \bibnamefont{Weiss}},
  \bibinfo{journal}{Nature} \textbf{\bibinfo{volume}{440}},
  \bibinfo{pages}{900} (\bibinfo{year}{2006}).

\bibitem[{\citenamefont{Hofferberth et~al.}(2007)\citenamefont{Hofferberth,
  Lesanovsky, Fischer, Schumm, and Schmiedmayer}}]{hofferberth_lesanovsky_07}
\bibinfo{author}{\bibfnamefont{S.}~\bibnamefont{Hofferberth}},
  \bibinfo{author}{\bibfnamefont{I.}~\bibnamefont{Lesanovsky}},
  \bibinfo{author}{\bibfnamefont{B.}~\bibnamefont{Fischer}},
  \bibinfo{author}{\bibfnamefont{T.}~\bibnamefont{Schumm}}, \bibnamefont{and}
  \bibinfo{author}{\bibfnamefont{J.}~\bibnamefont{Schmiedmayer}},
  \bibinfo{journal}{Nature} \textbf{\bibinfo{volume}{449}},
  \bibinfo{pages}{324} (\bibinfo{year}{2007}).

\bibitem[{\citenamefont{Hung et~al.}(2010)\citenamefont{Hung, Zhang, Gemelke,
  and Chin}}]{hung_zhang_10}
\bibinfo{author}{\bibfnamefont{C.-L.} \bibnamefont{Hung}},
  \bibinfo{author}{\bibfnamefont{X.}~\bibnamefont{Zhang}},
  \bibinfo{author}{\bibfnamefont{N.}~\bibnamefont{Gemelke}}, \bibnamefont{and}
  \bibinfo{author}{\bibfnamefont{C.}~\bibnamefont{Chin}},
  \bibinfo{journal}{Phys. Rev. Lett.} \textbf{\bibinfo{volume}{104}},
  \bibinfo{pages}{160403} (\bibinfo{year}{2010}).

\bibitem[{\citenamefont{{Trotzky} et~al.}(2011)\citenamefont{{Trotzky}, {C},
  {Flesch}, {McCulloch}, {Schollw{\"o}ck}, {Eisert}, and
  {Bloch}}}]{trotzky_chen_11}
\bibinfo{author}{\bibfnamefont{S.}~\bibnamefont{{Trotzky}}},
  \bibinfo{author}{\bibfnamefont{Y.-A.} \bibnamefont{{Chen}}},
  \bibinfo{author}{\bibfnamefont{A.}~\bibnamefont{{Flesch}}},
  \bibinfo{author}{\bibfnamefont{I.~P.} \bibnamefont{{McCulloch}}},
  \bibinfo{author}{\bibfnamefont{U.}~\bibnamefont{{Schollw{\"o}ck}}},
  \bibinfo{author}{\bibfnamefont{J.}~\bibnamefont{{Eisert}}}, \bibnamefont{and}
  \bibinfo{author}{\bibfnamefont{I.}~\bibnamefont{{Bloch}}},
  \bibinfo{journal}{Phys. Rev. Lett.} \textbf{\bibinfo{volume}{106}},
  \bibinfo{pages}{155302} (\bibinfo{year}{2011}).

\bibitem[{\citenamefont{Dziarmaga}(2010)}]{dziarmaga_10}
\bibinfo{author}{\bibfnamefont{J.}~\bibnamefont{Dziarmaga}},
  \bibinfo{journal}{Advances in Physics} \textbf{\bibinfo{volume}{59}},
  \bibinfo{pages}{1063} (\bibinfo{year}{2010}).

\bibitem[{\citenamefont{Cazalilla and Rigol}(2010)}]{cazalilla_rigol_10}
\bibinfo{author}{\bibfnamefont{M.~A.} \bibnamefont{Cazalilla}}
  \bibnamefont{and} \bibinfo{author}{\bibfnamefont{M.}~\bibnamefont{Rigol}},
  \bibinfo{journal}{New J. Phys.} \textbf{\bibinfo{volume}{12}},
  \bibinfo{pages}{055006} (\bibinfo{year}{2010}).

\bibitem[{\citenamefont{Polkovnikov et~al.}()\citenamefont{Polkovnikov,
  Sengupta, Silva, and Vengalattore}}]{polkovnikov_sengupta_11}
\bibinfo{author}{\bibfnamefont{A.}~\bibnamefont{Polkovnikov}},
  \bibinfo{author}{\bibfnamefont{K.}~\bibnamefont{Sengupta}},
  \bibinfo{author}{\bibfnamefont{A.}~\bibnamefont{Silva}}, \bibnamefont{and}
  \bibinfo{author}{\bibfnamefont{M.}~\bibnamefont{Vengalattore}},
\bibinfo{journal}{Rev. Mod. Phys.} \textbf{\bibinfo{volume}{83}},
  \bibinfo{pages}{863} (\bibinfo{year}{2011}).

\bibitem[{\citenamefont{Dalfovo et~al.}(1999)\citenamefont{Dalfovo, Giorgini,
  Pitaevskii, and Stringari}}]{dalfovo_giorgini_review_99}
\bibinfo{author}{\bibfnamefont{F.}~\bibnamefont{Dalfovo}},
  \bibinfo{author}{\bibfnamefont{S.}~\bibnamefont{Giorgini}},
  \bibinfo{author}{\bibfnamefont{L.~P.} \bibnamefont{Pitaevskii}},
  \bibnamefont{and}
  \bibinfo{author}{\bibfnamefont{S.}~\bibnamefont{Stringari}},
  \bibinfo{journal}{Rev. Mod. Phys.} \textbf{\bibinfo{volume}{71}},
  \bibinfo{pages}{463} (\bibinfo{year}{1999}).

\bibitem[{\citenamefont{Sutherland}(1998)}]{sutherland_98}
\bibinfo{author}{\bibfnamefont{B.}~\bibnamefont{Sutherland}},
  \bibinfo{journal}{Phys. Rev. Lett.} \textbf{\bibinfo{volume}{80}},
  \bibinfo{pages}{3678} (\bibinfo{year}{1998}).

\bibitem[{\citenamefont{Rigol and
  Muramatsu}(2005{\natexlab{a}})}]{rigol_muramatsu_05eHCBb}
\bibinfo{author}{\bibfnamefont{M.}~\bibnamefont{Rigol}} \bibnamefont{and}
  \bibinfo{author}{\bibfnamefont{A.}~\bibnamefont{Muramatsu}},
  \bibinfo{journal}{Phys. Rev. Lett.} \textbf{\bibinfo{volume}{94}},
  \bibinfo{pages}{240403} (\bibinfo{year}{2005}{\natexlab{a}}).

\bibitem[{\citenamefont{Minguzzi and Gangardt}(2005)}]{minguzzi_gangardt_05}
\bibinfo{author}{\bibfnamefont{A.}~\bibnamefont{Minguzzi}} \bibnamefont{and}
  \bibinfo{author}{\bibfnamefont{D.~M.} \bibnamefont{Gangardt}},
  \bibinfo{journal}{Phys. Rev. Lett.} \textbf{\bibinfo{volume}{94}},
  \bibinfo{pages}{240404} (\bibinfo{year}{2005}).

\bibitem[{\citenamefont{Heidrich-Meisner
  et~al.}(2009)\citenamefont{Heidrich-Meisner, Manmana, Rigol, Muramatsu,
  Feiguin, and Dagotto}}]{fabian_manmana_09}
\bibinfo{author}{\bibfnamefont{F.}~\bibnamefont{Heidrich-Meisner}},
  \bibinfo{author}{\bibfnamefont{S.~R.} \bibnamefont{Manmana}},
  \bibinfo{author}{\bibfnamefont{M.}~\bibnamefont{Rigol}},
  \bibinfo{author}{\bibfnamefont{A.}~\bibnamefont{Muramatsu}},
  \bibinfo{author}{\bibfnamefont{A.~E.} \bibnamefont{Feiguin}},
  \bibnamefont{and} \bibinfo{author}{\bibfnamefont{E.}~\bibnamefont{Dagotto}},
  \bibinfo{journal}{Phys. Rev. A} \textbf{\bibinfo{volume}{80}},
  \bibinfo{pages}{041603(R)} (\bibinfo{year}{2009}).

\bibitem[{\citenamefont{Fisher et~al.}(1989)\citenamefont{Fisher, Weichman,
  Grinstein, and Fisher}}]{fisher_weichman_89}
\bibinfo{author}{\bibfnamefont{M.~P.~A.} \bibnamefont{Fisher}},
  \bibinfo{author}{\bibfnamefont{P.~B.} \bibnamefont{Weichman}},
  \bibinfo{author}{\bibfnamefont{G.}~\bibnamefont{Grinstein}},
  \bibnamefont{and} \bibinfo{author}{\bibfnamefont{D.~S.}
  \bibnamefont{Fisher}}, \bibinfo{journal}{Phys. Rev. B}
  \textbf{\bibinfo{volume}{40}}, \bibinfo{pages}{546} (\bibinfo{year}{1989}).

\bibitem[{\citenamefont{Rigol and Muramatsu}(2004)}]{rigol_muramatsu_04eHBCa}
\bibinfo{author}{\bibfnamefont{M.}~\bibnamefont{Rigol}} \bibnamefont{and}
  \bibinfo{author}{\bibfnamefont{A.}~\bibnamefont{Muramatsu}},
  \bibinfo{journal}{Phys. Rev. Lett.} \textbf{\bibinfo{volume}{93}},
  \bibinfo{pages}{230404} (\bibinfo{year}{2004}).

\bibitem[{\citenamefont{Rigol and
  Muramatsu}(2005{\natexlab{b}})}]{rigol_muramatsu_05eHCBc}
\bibinfo{author}{\bibfnamefont{M.}~\bibnamefont{Rigol}} \bibnamefont{and}
  \bibinfo{author}{\bibfnamefont{A.}~\bibnamefont{Muramatsu}},
  \bibinfo{journal}{Mod. Phys. Lett.} \textbf{\bibinfo{volume}{19}},
  \bibinfo{pages}{861} (\bibinfo{year}{2005}{\natexlab{b}}).

\bibitem[{\citenamefont{Rodriguez et~al.}(2006)\citenamefont{Rodriguez,
  Manmana, Rigol, Noack, and Muramatsu}}]{rodriguez_manmana_06}
\bibinfo{author}{\bibfnamefont{K.}~\bibnamefont{Rodriguez}},
  \bibinfo{author}{\bibfnamefont{S.~R.} \bibnamefont{Manmana}},
  \bibinfo{author}{\bibfnamefont{M.}~\bibnamefont{Rigol}},
  \bibinfo{author}{\bibfnamefont{R.~M.} \bibnamefont{Noack}}, \bibnamefont{and}
  \bibinfo{author}{\bibfnamefont{A.}~\bibnamefont{Muramatsu}},
  \bibinfo{journal}{New J. Phys.} \textbf{\bibinfo{volume}{8}},
  \bibinfo{pages}{169} (\bibinfo{year}{2006}).

\bibitem[{\citenamefont{Lancaster and Mitra}(2010)}]{lancaster_mitra_10}
\bibinfo{author}{\bibfnamefont{J.}~\bibnamefont{Lancaster}} \bibnamefont{and}
  \bibinfo{author}{\bibfnamefont{A.}~\bibnamefont{Mitra}},
  \bibinfo{journal}{Phys. Rev. E} \textbf{\bibinfo{volume}{81}},
  \bibinfo{pages}{061134} (\bibinfo{year}{2010}).

\bibitem[{\citenamefont{Heidrich-Meisner
  et~al.}(2008)\citenamefont{Heidrich-Meisner, Rigol, Muramatsu, Feiguin, and
  Dagotto}}]{fabian_rigol_08}
\bibinfo{author}{\bibfnamefont{F.}~\bibnamefont{Heidrich-Meisner}},
  \bibinfo{author}{\bibfnamefont{M.}~\bibnamefont{Rigol}},
  \bibinfo{author}{\bibfnamefont{A.}~\bibnamefont{Muramatsu}},
  \bibinfo{author}{\bibfnamefont{A.~E.} \bibnamefont{Feiguin}},
  \bibnamefont{and} \bibinfo{author}{\bibfnamefont{E.}~\bibnamefont{Dagotto}},
  \bibinfo{journal}{Phys. Rev. A} \textbf{\bibinfo{volume}{78}},
  \bibinfo{pages}{013620} (\bibinfo{year}{2008}).

\bibitem[{\citenamefont{Schollw\"ock}(2005)}]{schollwock_review_05}
\bibinfo{author}{\bibfnamefont{U.}~\bibnamefont{Schollw\"ock}},
  \bibinfo{journal}{Rev. Mod. Phys.} \textbf{\bibinfo{volume}{77}},
  \bibinfo{pages}{259} (\bibinfo{year}{2005}).

\bibitem[{\citenamefont{Hen and Rigol}(2010)}]{hen_rigol_10}
\bibinfo{author}{\bibfnamefont{I.}~\bibnamefont{Hen}} \bibnamefont{and}
  \bibinfo{author}{\bibfnamefont{M.}~\bibnamefont{Rigol}},
  \bibinfo{journal}{Phys. Rev. Lett.} \textbf{\bibinfo{volume}{105}},
  \bibinfo{pages}{180401} (\bibinfo{year}{2010}).

\bibitem{Smith1} 
L.-K. Lim, C. M. Smith, and A. Hemmerich, 
Phys. Rev. Lett. {\bf 100}, 130402 (2008).

\bibitem{Smith2}
M. Di Liberto, O. Tieleman, V. Branchina, and C. Morais Smith,
Phys. Rev. A {\bf 84}, 013607 (2011).

\bibitem[{\citenamefont{Jaksch et~al.}(1998)\citenamefont{Jaksch, Bruder,
  Cirac, Gardiner, and Zoller}}]{jaksch_bruder_98}
\bibinfo{author}{\bibfnamefont{D.}~\bibnamefont{Jaksch}},
  \bibinfo{author}{\bibfnamefont{C.}~\bibnamefont{Bruder}},
  \bibinfo{author}{\bibfnamefont{J.~I.} \bibnamefont{Cirac}},
  \bibinfo{author}{\bibfnamefont{C.~W.} \bibnamefont{Gardiner}},
  \bibnamefont{and} \bibinfo{author}{\bibfnamefont{P.}~\bibnamefont{Zoller}},
  \bibinfo{journal}{Phys. Rev. Lett.} \textbf{\bibinfo{volume}{81}},
  \bibinfo{pages}{3108} (\bibinfo{year}{1998}).

\bibitem[{\citenamefont{Batrouni et~al.}(1990)\citenamefont{Batrouni,
  Scalettar, and Zimanyi}}]{batrouni_scalettar_90}
\bibinfo{author}{\bibfnamefont{G.~G.} \bibnamefont{Batrouni}},
  \bibinfo{author}{\bibfnamefont{R.~T.} \bibnamefont{Scalettar}},
  \bibnamefont{and} \bibinfo{author}{\bibfnamefont{G.~T.}
  \bibnamefont{Zimanyi}}, \bibinfo{journal}{Phys. Rev. Lett.}
  \textbf{\bibinfo{volume}{65}}, \bibinfo{pages}{1765} (\bibinfo{year}{1990}).

\bibitem[{\citenamefont{Freericks and Monien}(1996)}]{freericks_monien_96}
\bibinfo{author}{\bibfnamefont{J.~K.} \bibnamefont{Freericks}}
  \bibnamefont{and} \bibinfo{author}{\bibfnamefont{H.}~\bibnamefont{Monien}},
  \bibinfo{journal}{Phys. Rev. B} \textbf{\bibinfo{volume}{53}},
  \bibinfo{pages}{2691} (\bibinfo{year}{1996}).

\bibitem[{\citenamefont{Elstner and Monien}(1999)}]{elstner_monien_99}
\bibinfo{author}{\bibfnamefont{N.}~\bibnamefont{Elstner}} \bibnamefont{and}
  \bibinfo{author}{\bibfnamefont{H.}~\bibnamefont{Monien}},
  \bibinfo{journal}{Phys. Rev. B} \textbf{\bibinfo{volume}{59}},
  \bibinfo{pages}{12184} (\bibinfo{year}{1999}).

\bibitem[{\citenamefont{K\"uhner et~al.}(2000)\citenamefont{K\"uhner, White,
  and Monien}}]{kuhner_white_00}
\bibinfo{author}{\bibfnamefont{T.~D.} \bibnamefont{K\"uhner}},
  \bibinfo{author}{\bibfnamefont{S.~R.} \bibnamefont{White}}, \bibnamefont{and}
  \bibinfo{author}{\bibfnamefont{H.}~\bibnamefont{Monien}},
  \bibinfo{journal}{Phys. Rev. B} \textbf{\bibinfo{volume}{61}},
  \bibinfo{pages}{12474} (\bibinfo{year}{2000}).

\bibitem[{\citenamefont{Capogrosso-Sansone
  et~al.}(2007)\citenamefont{Capogrosso-Sansone, Prokof'ev, and
  Svistunov}}]{sansone_prokofev_07}
\bibinfo{author}{\bibfnamefont{B.}~\bibnamefont{Capogrosso-Sansone}},
  \bibinfo{author}{\bibfnamefont{N.~V.} \bibnamefont{Prokof'ev}},
  \bibnamefont{and} \bibinfo{author}{\bibfnamefont{B.~V.}
  \bibnamefont{Svistunov}}, \bibinfo{journal}{Phys. Rev. B}
  \textbf{\bibinfo{volume}{75}}, \bibinfo{pages}{134302}
  (\bibinfo{year}{2007}).

\bibitem[{\citenamefont{Capogrosso-Sansone
  et~al.}(2008)\citenamefont{Capogrosso-Sansone, S\"oyler, Prokof'ev, and
  Svistunov}}]{sansone_soyler_08}
\bibinfo{author}{\bibfnamefont{B.}~\bibnamefont{Capogrosso-Sansone}},
  \bibinfo{author}{\bibfnamefont{S.~G.} \bibnamefont{S\"oyler}},
  \bibinfo{author}{\bibfnamefont{N.~V.} \bibnamefont{Prokof'ev}},
  \bibnamefont{and} \bibinfo{author}{\bibfnamefont{B.~V.}
  \bibnamefont{Svistunov}}, \bibinfo{journal}{Phys. Rev. A}
  \textbf{\bibinfo{volume}{77}}, \bibinfo{pages}{015602}
  (\bibinfo{year}{2008}).

\bibitem[{\citenamefont{Batrouni et~al.}(2002)\citenamefont{Batrouni, Rousseau,
  Scalettar, Rigol, Muramatsu, Denteneer, and Troyer}}]{batrouni_rousseau_02}
\bibinfo{author}{\bibfnamefont{G.~G.} \bibnamefont{Batrouni}},
  \bibinfo{author}{\bibfnamefont{V.}~\bibnamefont{Rousseau}},
  \bibinfo{author}{\bibfnamefont{R.~T.} \bibnamefont{Scalettar}},
  \bibinfo{author}{\bibfnamefont{M.}~\bibnamefont{Rigol}},
  \bibinfo{author}{\bibfnamefont{A.}~\bibnamefont{Muramatsu}},
  \bibinfo{author}{\bibfnamefont{P.~J.~H.} \bibnamefont{Denteneer}},
  \bibnamefont{and} \bibinfo{author}{\bibfnamefont{M.}~\bibnamefont{Troyer}},
  \bibinfo{journal}{Phys. Rev. Lett.} \textbf{\bibinfo{volume}{89}},
  \bibinfo{pages}{117203} (\bibinfo{year}{2002}).

\bibitem[{\citenamefont{Wessel et~al.}(2004)\citenamefont{Wessel, Alet, Troyer,
  and Batrouni}}]{wessel_alet_04}
\bibinfo{author}{\bibfnamefont{S.}~\bibnamefont{Wessel}},
  \bibinfo{author}{\bibfnamefont{F.}~\bibnamefont{Alet}},
  \bibinfo{author}{\bibfnamefont{M.}~\bibnamefont{Troyer}}, \bibnamefont{and}
  \bibinfo{author}{\bibfnamefont{G.~G.} \bibnamefont{Batrouni}},
  \bibinfo{journal}{Phys. Rev. A} \textbf{\bibinfo{volume}{70}},
  \bibinfo{pages}{053615} (\bibinfo{year}{2004}).

\bibitem[{\citenamefont{Rigol et~al.}(2009)\citenamefont{Rigol, Batrouni,
  Rousseau, and Scalettar}}]{rigol_batrouni_09}
\bibinfo{author}{\bibfnamefont{M.}~\bibnamefont{Rigol}},
  \bibinfo{author}{\bibfnamefont{G.~G.} \bibnamefont{Batrouni}},
  \bibinfo{author}{\bibfnamefont{V.~G.} \bibnamefont{Rousseau}},
  \bibnamefont{and} \bibinfo{author}{\bibfnamefont{R.~T.}
  \bibnamefont{Scalettar}}, \bibinfo{journal}{Phys. Rev. A}
  \textbf{\bibinfo{volume}{79}}, \bibinfo{pages}{053605}
  (\bibinfo{year}{2009}).

\bibitem[{\citenamefont{F\"olling et~al.}(2006)\citenamefont{F\"olling, Widera,
  M\"uller, Gerbier, and Bloch}}]{folling_widera_06}
\bibinfo{author}{\bibfnamefont{S.}~\bibnamefont{F\"olling}},
  \bibinfo{author}{\bibfnamefont{A.}~\bibnamefont{Widera}},
  \bibinfo{author}{\bibfnamefont{T.}~\bibnamefont{M\"uller}},
  \bibinfo{author}{\bibfnamefont{F.}~\bibnamefont{Gerbier}}, \bibnamefont{and}
  \bibinfo{author}{\bibfnamefont{I.}~\bibnamefont{Bloch}},
  \bibinfo{journal}{Phys. Rev. Lett.} \textbf{\bibinfo{volume}{97}},
  \bibinfo{pages}{060403} (\bibinfo{year}{2006}).

\bibitem[{\citenamefont{Campbell et~al.}(2006)\citenamefont{Campbell, Mun,
  Boyd, Medley, Leanhardt, Marcassa, Pritchard, and
  Ketterle}}]{campbell_mun_06}
\bibinfo{author}{\bibfnamefont{G.~K.} \bibnamefont{Campbell}},
  \bibinfo{author}{\bibfnamefont{J.}~\bibnamefont{Mun}},
  \bibinfo{author}{\bibfnamefont{M.}~\bibnamefont{Boyd}},
  \bibinfo{author}{\bibfnamefont{P.}~\bibnamefont{Medley}},
  \bibinfo{author}{\bibfnamefont{A.~E.} \bibnamefont{Leanhardt}},
  \bibinfo{author}{\bibfnamefont{L.~G.} \bibnamefont{Marcassa}},
  \bibinfo{author}{\bibfnamefont{D.~E.} \bibnamefont{Pritchard}},
  \bibnamefont{and} \bibinfo{author}{\bibfnamefont{W.}~\bibnamefont{Ketterle}},
  \bibinfo{journal}{Science} \textbf{\bibinfo{volume}{313}},
  \bibinfo{pages}{649} (\bibinfo{year}{2006}).

\bibitem[{\citenamefont{Gemelke et~al.}(2009)\citenamefont{Gemelke, Zhang,
  Hung, and Chin}}]{gemelke_Zhang_09}
\bibinfo{author}{\bibfnamefont{N.}~\bibnamefont{Gemelke}},
  \bibinfo{author}{\bibfnamefont{X.}~\bibnamefont{Zhang}},
  \bibinfo{author}{\bibfnamefont{C.-L.} \bibnamefont{Hung}}, \bibnamefont{and}
  \bibinfo{author}{\bibfnamefont{C.}~\bibnamefont{Chin}},
  \bibinfo{journal}{Nature} \textbf{\bibinfo{volume}{460}},
  \bibinfo{pages}{995} (\bibinfo{year}{2009}).

\bibitem[{\citenamefont{Bakr et~al.}(2010)\citenamefont{Bakr, Peng, Tai, Ma,
  Simon, Gillen, Folling, Pollet, and Greiner}}]{bakr_peng_10}
\bibinfo{author}{\bibfnamefont{W.~S.} \bibnamefont{Bakr}},
  \bibinfo{author}{\bibfnamefont{A.}~\bibnamefont{Peng}},
  \bibinfo{author}{\bibfnamefont{M.~E.} \bibnamefont{Tai}},
  \bibinfo{author}{\bibfnamefont{R.}~\bibnamefont{Ma}},
  \bibinfo{author}{\bibfnamefont{J.}~\bibnamefont{Simon}},
  \bibinfo{author}{\bibfnamefont{J.~I.} \bibnamefont{Gillen}},
  \bibinfo{author}{\bibfnamefont{S.}~\bibnamefont{Folling}},
  \bibinfo{author}{\bibfnamefont{L.}~\bibnamefont{Pollet}}, \bibnamefont{and}
  \bibinfo{author}{\bibfnamefont{M.}~\bibnamefont{Greiner}},
  \bibinfo{journal}{Science} \textbf{\bibinfo{volume}{329}},
  \bibinfo{pages}{547} (\bibinfo{year}{2010}).

\bibitem[{\citenamefont{Sherson et~al.}(2010)\citenamefont{Sherson, Weitenberg,
  Endres, Cheneau, Bloch, and Kuhr}}]{sherson_weitenberg_10}
\bibinfo{author}{\bibfnamefont{J.~F.} \bibnamefont{Sherson}},
  \bibinfo{author}{\bibfnamefont{C.}~\bibnamefont{Weitenberg}},
  \bibinfo{author}{\bibfnamefont{M.}~\bibnamefont{Endres}},
  \bibinfo{author}{\bibfnamefont{M.}~\bibnamefont{Cheneau}},
  \bibinfo{author}{\bibfnamefont{I.}~\bibnamefont{Bloch}}, \bibnamefont{and}
  \bibinfo{author}{\bibfnamefont{S.}~\bibnamefont{Kuhr}},
  \bibinfo{journal}{Nature} \textbf{\bibinfo{volume}{467}}, \bibinfo{pages}{68}
  (\bibinfo{year}{2010}).

\bibitem[{\citenamefont{Sheshadri}(1993)}]{sheshadri_Krishnamurthy_93}
\bibinfo{author}{\bibfnamefont{K.}~\bibnamefont{Sheshadri}},
\bibinfo{author}{\bibfnamefont{H.R.}~\bibnamefont{Krishnamurthy}},
\bibinfo{author}{\bibfnamefont{R.}~\bibnamefont{Pandit}},
\bibinfo{author}{\bibfnamefont{T.V.}~\bibnamefont{Ramakrishnan}},
\bibinfo{journal}{Europhys. Lett.} \textbf{\bibinfo{volume}{22}},
\bibinfo{pages}{257} (\bibinfo{year}{1993}).


\bibitem[{\citenamefont{Amico and Penna}(1998)}]{amico_penna_98}
\bibinfo{author}{\bibfnamefont{L.}~\bibnamefont{Amico}} \bibnamefont{and}
  \bibinfo{author}{\bibfnamefont{V.}~\bibnamefont{Penna}},
  \bibinfo{journal}{Phys. Rev. Lett.} \textbf{\bibinfo{volume}{80}},
  \bibinfo{pages}{2189} (\bibinfo{year}{1998}).

\bibitem[{\citenamefont{Jaksch et~al.}(2002)\citenamefont{Jaksch, Venturi,
  Cirac, Williams, and Zoller}}]{jaksch_venturi_02}
\bibinfo{author}{\bibfnamefont{D.}~\bibnamefont{Jaksch}},
  \bibinfo{author}{\bibfnamefont{V.}~\bibnamefont{Venturi}},
  \bibinfo{author}{\bibfnamefont{J.~I.} \bibnamefont{Cirac}},
  \bibinfo{author}{\bibfnamefont{C.~J.} \bibnamefont{Williams}},
  \bibnamefont{and} \bibinfo{author}{\bibfnamefont{P.}~\bibnamefont{Zoller}},
  \bibinfo{journal}{Phys. Rev. Lett.} \textbf{\bibinfo{volume}{89}},
  \bibinfo{pages}{040402} (\bibinfo{year}{2002}).

\bibitem[{\citenamefont{Zakrzewski}(2005)}]{zakrzewski_01}
\bibinfo{author}{\bibfnamefont{J.}~\bibnamefont{Zakrzewski}},
  \bibinfo{journal}{Phys. Rev. A} \textbf{\bibinfo{volume}{71}},
  \bibinfo{pages}{043601} (\bibinfo{year}{2005}).


\bibitem[{\citenamefont{Snoek}(2011)}]{snoek}
\bibinfo{author}{\bibfnamefont{M.}~\bibnamefont{Snoek}},
  \bibinfo{journal}{Europhys. Lett.} \textbf{\bibinfo{volume}{95}},
  \bibinfo{pages}{30006} (\bibinfo{year}{2011}).


\bibitem[{\citenamefont{Sachdev}(2000)}]{sachdev_00}
\bibinfo{author}{\bibfnamefont{S.}~\bibnamefont{Sachdev}},
\bibinfo{title}{Quantum Phase Transitions,} \bibinfo{publisher}{Cambridge University
 Press}, \bibinfo{year}{2000}.

\bibitem{penrose56} O. Penrose and L. Onsager, Phys. Rev. {\bf 104}, 
576 (1956).

\bibitem{leggett01}
A. J. Leggett, Rev. Mod. Phys. {\bf 73}, 307 (2001).

\bibitem[{\citenamefont{buchhold}(2011)}]{buchhold_11}
\bibinfo{author}{\bibfnamefont{M.}~\bibnamefont{Buchhold}},
\bibinfo{author}{\bibfnamefont{U.}~\bibnamefont{Bissbort}},
\bibinfo{author}{\bibfnamefont{S.}~\bibnamefont{Will}},
\bibnamefont{and} \bibinfo{author}{\bibfnamefont{W.}~\bibnamefont{Hofstetter}},
\bibinfo{journal}{Phys. Rev. A} \textbf{\bibinfo{volume}{84}},
\bibinfo{pages}{023631} (\bibinfo{year}{2011}).


\end{thebibliography}
\end{document}